\documentclass[12pt]{article}
\usepackage{graphicx}
\begin{document}

\def\Nh{\hat N}
\def\Mh{\hat M}
\def\Bh{\hat B}
\def\Bm{B^{(m)}}
\def\Bhm{{\hat B}^{(m)}}
\def\Bht{\hat {\tilde B}}
\def\Bhtm{{\hat {\tilde B}}_{(m)}}
\def\ft{\tilde f}
\def\gt{\tilde g}
\def\bt{\tilde b}
\def\qt{\tilde q}
\def\pt{\tilde p}
\def\Bt{\tilde B}
\def\Btm{{\tilde B}_{(m)}}
\def\Qt{\tilde Q}
\def\Nt{\tilde N}
\def\SN{\alpha}

\begin{titlepage}
\thispagestyle{empty}
\begin{flushright}
BROWN-HET-1353
\end{flushright}
\vskip 1cm
\begin{center}
{\LARGE\bf Notes on anomalies, baryons, and Seiberg duality}
\vskip 1cm

{\large Steven Corley }
\vskip .5cm
{\it Department of Physics \\
Brown  University \\
Providence, RI 02912}\\
{\tt scorley@het.brown.edu}

\end{center}
\vskip 1cm

\begin{abstract}
We consider an ${\cal N}=1$ $SU(N_c)$ SUSY gauge theory
with $N_f \geq N_c$ matter multiplets
transforming in the fundamental and antifundamental
representations of the gauge group.  Using the Konishi anomaly 
and a non-anomalous conservation law, we derive
a system of partial differential equations that
determine the low energy effective superpotential
as a function of the mesonic and baryonic vacuum 
expectation values.
We apply the formalism to the cases of $N_f = N_c$ and 
$N_f = N_c +1$ where the equations are easily
integrated and recover the known results.  We further
apply the formalism to derive a system of partial
differential equations to determine the low energy
effective superpotential for the Seiberg dual 
theories. Finally we briefly discuss the
associated matrix models via the Dijkgraaf-Vafa
conjecture.
\end{abstract}

\end{titlepage}

\section{Introduction}

Quantum field theories with ${\cal N} = 1$ supersymmetry
were studied in great detail in the mid 90's, for reviews
see eg. \cite{Intriligator:1996au,Peskin:1997qi}.  One of the 
problems addressed was that
of the dynamically generated superpotential.  Seiberg
\cite{Seiberg:1993vc} showed that one can go a long way in determining
the exact form of the superpotential for various
theories using symmetries and holomorphy. 

Recently a new technique for evaluating low energy effective
superpotentials has been developed by Dijkgraaf and Vafa 
\cite{Dijkgraaf:2002fc,Dijkgraaf:2002vw,Dijkgraaf:2002dh}.
They conjectured that the exact low energy effective superpotential
for an ${\cal N}=1$ SUSY gauge theory with an adjoint matter
field could be constructed from an associated matrix model.
The action of the matrix model consists of the superpotential
appearing in the definition of the gauge theory and the 
variables of the matrix model are the superfields themselves.
The conjecture then relates the planar diagrammatric contribution
to the free energy of the matrix model to the exact low energy
effective superpotential of the corresponding gauge theory.

Three different proofs of the conjecture have been given,
\cite{Ferrari:2002jp,Dijkgraaf:2002xd,Cachazo:2002ry}.
\cite{Ferrari:2002jp} treated the theory as an
${\cal N}=2$ theory broken to ${\cal N}=1$ by
the presence of a superpotential and then analyzed the gauge
theory using the Seiberg-Witten formalism
\cite{Seiberg:1994rs,Seiberg:1994aj}.  The low energy
effective superpotential was constructed and shown to
be the same as arising from the associated matrix model
via the Dijkgraaf-Vafa conjecture.
In \cite{Dijkgraaf:2002xd} the gauge theory was analyzed perturbatively
with the result that the diagrams relevant to the low
energy effective superpotential were indeed planar
and corresponded directly to planar graphs of the
associated matrix model.  In \cite{Cachazo:2002ry} a generalized
Konishi anomaly equation was constructed which takes the
exact form of a matrix model loop equation.  The
solution of the anomaly equations was shown to 
correspond to the solution of the associated
matrix model loop equations in exactly the way
conjectured by Dijkgraaf and Vafa.

The purpose of this paper is to analyze the
${\cal N}=1$ SUSY gauge theory with fundamental
matter only along
the lines of the Konishi
anomaly analysis of \cite{Cachazo:2002ry}.
This particular topic has been studied
in \cite{Seiberg:2002jq,Feng:2002is, 
Brandhuber:2003va,Balasubramanian:2003tv,Cachazo:2003yc,Kraus:2003jv}.
Specifically we
shall be interested in theories with the number
of flavors satisfying $N_f \geq N_c$ and
vanishing superpotential.  For such theories
the low energy degrees of freedom consists
of mesons and baryons.  For $N_f = N_c$
and $N_f = N_c + 1$ the exact results are
known, \cite{Seiberg:1994bz}.  In the former case there is 
no dynamically generated 
superpotential but the classical constraint
equation on the meson and baryon vacuum 
expectation values is modified.  In the
latter case there is a dynamically generated
superpotential which serves to enforce the
classical constraint equations on the
meson and baryon vevs.

The remaining cases $N_f \geq N_c + 2$ are less
well understood.  For $3 N_c < N_f$ the theory is
infrared free while 
for $3/2 N_c \leq N_f \leq 3 N_c$ the theory is
believed to be a conformal field theory.
In the range $N_c + 1 < N_f < 3/2 N_c$ the
theory can no longer be conformal as such
an assumption would violate unitarity.
Instead Seiberg has conjectured \cite{Seiberg:1995pq} that
the theory has a dual description in
terms of an $SU(N_f - N_c)$ gauge theory
with $N_f$ fundamental matter multiplets along with
a set of scalar multiplets.  This conjecture
has passed various consistency checks including
showing that both theories have the same
global symmetries, have matching quantum numbers
for the gauge invariant operators, and have matching
anomalies.  

By deriving the Konishi anomaly equation, along with
a non-anomalous conservation law, we construct
a set of partial differential equations for
the low energy effective superpotential $W_{eff}$
as a function of the meson and baryon operators
for $N_f \geq N_c$.  The elementary superpotential
of the theory consists of terms giving vevs
to the mesons and baryons of the theory enforced
by introducing Lagrange multiplier fields.
The Konishi anomaly and conservation equations
are enough to solve for the Lagrange mulipliers
in terms of the meson and baryon operators.
These solutions are then used to construct a
set of partial differential equations for $W_{eff}$
determining it up to a function of the glueball
superfield.

In sections 2 and 3 we present the derivations
and results for the $N_f = N_c$ and $N_f = N_c + 1$
cases respectively.  In these cases we can solve
the equations completely and are able to recover
the results of \cite{Seiberg:1994bz}.  In sections 4 and 5 we
derive the Konishi anomaly and conservation equations
for the theory and its Seiberg dual respectively
for $N_f \geq N_c + 2$.  In this case the equations
become considerably more complicated and we are
unable to solve them.  Nevertheless the set of 
equations could in principle be used to provide
a proof of Seiberg's conjectured duality.  In
section 5 we consider the associated matrix model
and derive its loop equations.

\section{$N_f = N_c$}

The situation with $N_f = N_c = N$ is somewhat simpler, so we
begin with it.  As our superpotential we shall simply
take
\begin{equation}
W = m^{\ft}_{\, f} ( \Mh^{f}_{\, \ft} - M^{f}_{\, \ft})
+ b ( \Bh - B) + \bt (\Bht - \Bt)
\label{superzero}
\end{equation}
where we have defined the meson and baryon operators
\begin{eqnarray}
\Mh^{f}_{\, \ft} & := & \Qt_{\ft}^a Q^{f}_{a} \\
\Bh & := & \det Q = \frac{1}{N !} \epsilon_{f_1 \cdots f_N}
\epsilon^{a_1 \cdots a_N} Q^{f_1}_{a_1} \cdots Q^{f_N}_{a_N} \\
\Bht & := & \det \Qt = \frac{1}{N !} \epsilon^{\ft_1 \cdots \ft_N}
\epsilon_{a_1 \cdots a_N} \Qt_{\ft_1}^{a_1} \cdots \Qt_{\ft_N}^{a_N}.
\end{eqnarray}
The fields $m^{\ft}_{\, f}$, $b$, and $\bt$ are Lagrange 
multipliers that enforce the conditions that the
vacuum expectation values of the meson and baryon fields
are given respectively by $M^{f}_{\, \ft}$, $B$ and $\Bt$.
In the context of matrix models and the Dijkgraaf-Vafa
conjecture, this kind of superpotential was introduced
in \cite{Demasure:2002sc} for the mesons and generalized
to include baryons in \cite{Bena:2002ua}.
This theory has an $SU (N_f)_Q \times SU (N_f)_{\Qt}
\times U(1)_R \times \prod_{i=1}^{N_f} U(1)_{i,Q} \times 
\prod_{i=1}^{N_f} U(1)_{i,\Qt}$ symmetry\footnote{Note that the
$U(1)$ flavor symmetries are not all independent because
of the $SU(N_f)$ symmetries.} with
the transformation properties of the fields given in the
table below.  

\begin{tabular}{c|ccccc}
 & $SU (N_f)_Q$ & $SU (N_f)_{\Qt}$ & $U(1)_{i,Q}$ & $U(1)_{i,\Qt}$
& $U(1)_R$ \\ \hline
$Q^{f}:$ & $N_f$ & 1 & $\delta^{f}_{i}$ & 0 & $(N_f - N_c)/N_f$ \\
$\Qt_{\ft}:$ & 1 & ${\bar N}_f$ & 0 & $\delta^{i}_{\ft}$ & $(N_f - N_c)/N_f$ \\
\end{tabular}

\vspace{1em}
If there is no superpotential then the flavor symmetries
would classically give rise to the conservation equations
${\bar D}^2 (Q^{\dagger}_{f}
e^V Q^{f'}) = 0$ and $D^2 (\Qt^{\ft \dagger} e^V \Qt_{\ft'}) = 0$.
The $U(1)$ symmetries are however anomalous and 
are modified at one-loop \cite{Konishi:1984hf,Konishi:1985tu} to 
\begin{eqnarray}
{\bar D}^2 (Q^{\dagger}_{f'} e^V Q^{f}) & = &  S \delta^{f}_{f'} \\
{\bar D}^2 (\Qt_{\ft} e^V \Qt^{\ft' \dagger}) & = &  S \delta^{\ft'}_{\ft}
\label{anomtempf}
\end{eqnarray}
respectively, where the glueball superfield $S$ is
defined as
\begin{equation}
S := \frac{1}{32 \pi^2} Tr W^{\alpha} W_{\alpha}.
\end{equation}
Reinstating the superpotential (\ref{superzero}), the anomaly
equations pick up some extra terms from the equations of
motion for $Q^{\dagger}_{f'}$ and $\Qt^{\ft' \dagger}$ and become
\begin{eqnarray}
{\bar D}^2 (Q^{\dagger}_{f'} e^V Q^{f}) & = &  S \delta^{f}_{f'}
+ \Mh^{f}_{\, \ft} m^{\ft}_{\, f'} + b \Bh \delta^{f}_{f'} \nonumber \\
{\bar D}^2 (\Qt_{\ft} e^V \Qt^{\ft' \dagger}) & = &  S \delta^{\ft'}_{\ft}
+ m^{\ft'}_{\, f} \Mh^{f}_{\, \ft} + \bt \Bht \delta^{\ft'}_{\ft}
\label{anomtemp}
\end{eqnarray}
respectively.  Taking expectation values in a supersymmetric
vacuum will remove the left-hand-sides of these equations (or
equivalently working in the chiral ring of operators modded
out by ${\bar D}$ exact operators).  We finally arrive at
\begin{eqnarray}
\langle \Mh^{f}_{\, \ft} \rangle m^{\ft}_{\, f'} & = & 
( \langle S \rangle - b \langle \Bh \rangle ) \delta^{f}_{f'} \nonumber \\
m^{\ft'}_{\, f} \langle \Mh^{f}_{\, \ft} \rangle  & = & 
( \langle S \rangle - \bt \langle \Bht \rangle) \delta^{\ft'}_{\ft}.
\label{first}
\end{eqnarray}

We can obtain another set of equations similar to the above
set by noting that there is another, gauge invariant, conserved 
current.  Define the pair of operators
\begin{eqnarray}
\Bh_{f_1}^{a_1} & := & \frac{1}{(N-1)!} \epsilon^{a_1 \cdots a_N}
\epsilon_{f_1 \cdots f_N} Q^{f_2}_{a_2} \cdots Q^{f_N}_{a_N} 
= \Bh (Q^{-1})^{a_1}_{f_1} \nonumber \\
\Bht^{\ft_1}_{a_1} & := & \frac{1}{(N-1)!} \epsilon_{a_1 \cdots a_N}
\epsilon^{\ft_1 \cdots \ft_N} \Qt_{\ft_2}^{a_2} \cdots \Qt_{\ft_N}^{a_N} 
= \Bht (\Qt^{-1})_{a_1}^{\ft_1}.
\end{eqnarray}
Under $SU(N_c)$ these transform exactly as $\Qt_{\ft}^a$ and $Q^{f}_a$ 
respectively and therefore in the classical conservation laws
above we may replace $\Qt_{\ft}^a$ by $\Bh_{f}^a$ and 
$Q^{f}_a$ by $\Bht^{\ft}_a$ to obtain new conservation 
laws\footnote{Such generalizations have also been considered
in \cite{Brandhuber:2003va}}.
Unlike the cases above however, these new conservation laws
are not anomalous (as there is no propagator for $Q^{\dagger}$
and $\Qt$ or for $\Qt^{\dagger}$ and $Q$).  Including the
terms arising from the superpotential we obtain
\begin{eqnarray}
{\bar D}^2 (Q^{\dagger}_{f'} e^V \Bht^{\ft}) & = &
m^{\ft}_{\, f} \Bht + b \Bh \Bht (\Mh^{-1})^{\ft}_{\, f} \nonumber \\
{\bar D}^2 (\Bh_{f} e^V \Qt^{\ft' \dagger}) & = &
m^{\ft}_{\, f} \Bh + \bt \Bh \Bht (\Mh^{-1})^{\ft}_{\, f}
\label{secondinter}
\end{eqnarray}
which may equivalently be written as
\begin{eqnarray}
{\bar D}^2 (Q^{\dagger}_{f'} e^V \Bht^{\ft}) & = &
m^{\ft}_{\, f} \Bht + b \det \Mh  (\Mh^{-1})^{\ft}_{\, f} \nonumber \\
{\bar D}^2 (\Bh_{f} e^V \Qt^{\ft' \dagger}) & = &
m^{\ft}_{\, f} \Bh + \bt \det \Mh (\Mh^{-1})^{\ft}_{\, f}
\label{secondset}
\end{eqnarray}
where now the determinant is taken over the flavor indices.
Taking expectation values as before removes the left-hand-side.

It was shown in \cite{Cachazo:2002ry} that an expectation value
of a product of gauge invariant, chiral superfields is equal
to the product of the individual expectation values.  
Applying this to the expectation values of (\ref{secondset})
results in the equations
\begin{eqnarray}
m^{\ft}_{\, f} \langle \Bht \rangle + 
b \det \langle \Mh \rangle   (\langle \Mh \rangle^{-1})^{\ft}_{\, f}
& = & 0 \nonumber \\
m^{\ft}_{\, f} \langle \Bh \rangle + 
\bt \det \langle \Mh \rangle (\langle \Mh \rangle^{-1})^{\ft}_{\, f}
& = & 0.
\label{second}
\end{eqnarray}
We must be careful in obtaining this result.  We could have
taken the expectation values of these same equations in the
form of (\ref{secondinter}).  Applying the factorization
property of \cite{Cachazo:2002ry} would have resulted
(after some trivial manipulations) in the same equations as
in (\ref{first}) althought without the $S$ dependent term.
Taken together these two sets of equations would have implied
that $S$ must vanish, which is not correct.

Our goal now is to use these equations in order to solve
for the low energy effective action $W_{eff} (S,M,B,\Bt)$.
To do this we note that\footnote{To simplify the notation
we shall not write the expectation value signs $\langle \cdots \rangle$
from now on.}
\begin{eqnarray}
\frac{\partial W_{eff}}{\partial M^{f}_{\, \ft}} & = & 
- m^{\ft}_{\, f} \nonumber \\
\frac{\partial W_{eff}}{\partial B} & = & - b \nonumber \\
\frac{\partial W_{eff}}{\partial \Bt} & = & - \bt.
\label{dWzero}
\end{eqnarray}
Therefore if we can solve the above equations for the
Lagrange multiplier fields in terms of the variables 
appearing in the effective action, then we can integrate
to find the effective action up to an $S$ dependent term.

To begin we note that (\ref{first}) implies that
\begin{equation}
b \Bh = \bt \Bht
\end{equation}
and therefore that
\begin{equation}
\Bh = \bt \xi, \,\, \Bht = b \xi
\label{Btemp}
\end{equation}
where we have introduced a new variable $\xi$.
We may therefore rewrite (\ref{first}) as
\begin{equation}
\Mh^{f}_{\, \ft} = ( S - \frac{\Bh \Bht}{\xi} ) (m^{-1})^{f}_{\, \ft}.
\label{Mtemp}
\end{equation}
We have at this stage reduced the problem to solving for
$\xi$ in terms of the meson and baryon vevs.

To find an equation for $\xi$ we need only eliminate the Lagrange
multipliers $m^{\ft}_{\, f}$ and $b$ (or $\bt$) from (\ref{second})
using the expressions in (\ref{Mtemp}) and (\ref{Btemp}).  This
results in a linear equation for $\xi$ with solution
\begin{equation}
\xi = - \frac{\det \Mh - \Bh \Bht}{ S}.
\end{equation}
Substituting for $\xi$ in (\ref{Btemp}) and (\ref{Mtemp}) we find
\begin{eqnarray}
b & = & -  S \frac{\Bt}{\det M - B \Bt} \nonumber \\
\bt & = & -  S \frac{B}{\det M - B \Bt} \nonumber \\
m^{f}_{\, \ft} & = &  S \frac{ \det M}{\det M - B \Bt} 
(M^{-1})^{f}_{\, \ft}.
\end{eqnarray}
Finally we can insert these expressions back into the
derivative equations (\ref{dWzero}) for $W_{eff}$ and
integrate.  This results in
\begin{equation}
W_{eff} = - S \ln \Bigl( \frac{\det M - B \Bt}{\Lambda^{2 N}} \Bigr) + f(S)
\end{equation}
where $f(S)$ remains to be fixed and we have put in the scale
$\Lambda$ in order to make the argument of the logarithm
dimensionless.  $f(S)$ is just the Veneziano-Yankielowicz contribution
\begin{equation}
W_{VY}(S) = (N_c - N_f) S \Bigl( - \ln \frac{S}{\Lambda^3} + 1 \Bigr)
\label{WVY}
\end{equation}
which vanishes in this case as $N_c = N_f$.  Therefore setting
$f(S)$ to zero and minimizing $W_{eff}$ with respect to $S$
results in the constraint
\begin{equation}
\det M - B \Bt = \Lambda^{2 N}
\label{modconstraint}
\end{equation}
and moreover a vanishing superpotential.  This is in agreement
with the results of \cite{Seiberg:1994bz}.

\section{$N_f$ = $N_c$ + 1}

The methodology in the case $N_f = N_c + 1$ is identical to the
previous section, although the details are slightly more
involved.  As before the superpotential that we consider 
consists of Lagrange multiplier terms that fix the
vevs of the gauge invariant variables, i.e.,
\begin{equation}
W = m^{\ft}_{\, f} ( \Mh^{f}_{\, \ft} - M^{f}_{\, \ft})
+ b^f ( \Bh_f - B_f) + \bt_{\ft} (\Bht^{\ft} - \Bt^{\ft})
\label{superone}
\end{equation}
where the meson operator was defined before, the
baryon operators are defined as 
\begin{eqnarray}
\Bh_f & := &  \frac{1}{N !} \epsilon_{f f_1 \cdots f_N}
\epsilon^{a_1 \cdots a_N} Q^{f_1}_{a_1} \cdots Q^{f_N}_{a_N} \\
\Bht^{\ft} & := & \frac{1}{N !} \epsilon^{\ft \ft_1 \cdots \ft_N}
\epsilon_{a_1 \cdots a_N} \Qt_{\ft_1}^{a_1} \cdots \Qt_{\ft_N}^{a_N}
\end{eqnarray}
and the fields $m^{\ft}_{\, f}$, $b^f$, and $\bt_{\ft}$ are Lagrange
mulitpliers.

Proceeding exactly as before, we must derive the ``loop'' equations
analogous to those found in (\ref{first}) and (\ref{second}).
In the case of (\ref{first}) the steps are virtually identical,
so we shall forego the details and simply record the result.
After taking expectation values we find
\begin{eqnarray}
\langle \Mh^{f}_{\, \ft} \rangle m^{\ft}_{\, f'} & = & 
( \langle S \rangle - b^g \langle \Bh_g \rangle ) \delta^{f}_{f'}
+ b^f \langle \Bh_{f'} \rangle \nonumber \\
m^{\ft'}_{\, f} \langle \Mh^{f}_{\, \ft} \rangle  & = & 
( \langle S \rangle - \bt_{\gt} \langle \Bht^{\gt} \rangle) 
\delta^{\ft'}_{\ft} + \bt_{\ft} \langle \Bht^{\ft'} \rangle.
\label{first+1}
\end{eqnarray}

The second set of equations follows analogously to those
found in (\ref{second}).  Namely we note that 
$Q_{a}^{f}$ transforms under $SU(N_c)$ in the same
way as the operator
\begin{equation}
\Bht_{a}^{\ft \ft'} := \frac{1}{N!} \epsilon^{\ft \ft' \ft_2 \cdots \ft_N}
\epsilon_{a a_2 \cdots a_N} \Qt_{\ft_2}^{a_2} \cdots \Qt_{f_N}^{a_N}
\end{equation}
and therefore for vanishing superpotential we may 
define a new conserved current
$(Q^{\dagger}_{f'} e^V \Bht^{\ft \ft'})$, which is not anomalous.
In a completely analogous way we note that $\Qt^{a}_{\ft}$
transforms under color in the same way as
\begin{equation}
\Bh^{a}_{f f'} := \frac{1}{N!} \epsilon_{f f' f_2 \cdots f_N}
\epsilon^{a a_2 \cdots a_N} Q^{f_2}_{a_2} \cdots Q^{f_N}_{a_N}
\end{equation}
and therefore for vanishing superpotential one has 
the conserved current $(\Bh_{f f'} e^V \Qt^{\ft' \dagger})$.
Reinstating the superpotential (\ref{superone}) and taking
expectation values results in the equations
\begin{eqnarray}
(N-1)! (m^{\ft}_{\, f} \langle \Bht^{\ft'} \rangle - 
m^{\ft'}_{\, f} \langle \Bht^{\ft} \rangle) 
+ b^g \epsilon_{f_1 \cdots f_{N-1} f g} 
\epsilon^{\ft_1 \cdots \ft_{N-1} \ft \ft'} 
\langle \Mh^{f_1}_{\, \ft_1} \rangle \cdots 
\langle \Mh^{f_{N-1}}_{\, \ft_{N-1}} \rangle & = & 0 
\nonumber \\
(N-1)! (m^{\ft}_{\, f} \langle \Bh_{f'} \rangle - 
m^{\ft}_{\, f'} \langle \Bh_{f} \rangle )
+ \bt_{\gt} \epsilon^{\ft_1 \cdots \ft_{N-1} \ft \gt} 
\epsilon_{f_1 \cdots f_{N-1} f f'} 
\langle \Mh^{f_1}_{\, \ft_1} \rangle  \cdots 
\langle \Mh^{f_{N-1}}_{\, \ft_{N-1}} \rangle & = & 0 
\label{second+1}
\end{eqnarray}
where as before we have used the factorization property
of \cite{Cachazo:2002ry}.

Now we can proceed to solve these equations.  Recall that
the goal is to construct the low energy effective action
$W_{eff}(S,M,B_f,\Bt^{\ft})$ where $W_{eff}$ satisfies
the partial differential equations
\begin{eqnarray}
\frac{\partial W_{eff}}{\partial M^{f}_{\, \ft}} & = & 
- m^{\ft}_{\, f} \nonumber \\
\frac{\partial W_{eff}}{\partial B_f} & = & - b^f \nonumber \\
\frac{\partial W_{eff}}{\partial \Bt^{\ft}} & = & - \bt_{\ft}.
\label{dWone}
\end{eqnarray}
We can use the equations (\ref{first+1}) and (\ref{second+1})
to solve for $m^{\ft}_{\, f}$, $b^f$ and $\bt_{\ft}$ in terms
of $S$, $M^{f}_{\, \ft}$, $B_f$ and $\Bt^{\ft}$, and then 
integrate to find $W_{eff}$ up to an $S$ dependent term.

To proceed let's first rewrite (\ref{first+1}) in the
form
\begin{eqnarray}
\Mh^{f}_{\, \ft}  & = & 
( S - b^g \Bh_g ) (m^{-1})^{f}_{\ft}
+ b^f \Bh_{f'} (m^{-1})^{f'}_{\ft} \nonumber \\
\Mh^{f}_{\, \ft} & = & 
( S  - \bt_{\gt} \Bht^{\gt}) 
(m^{-1})^{f}_{\ft} + \bt_{\ft} (m^{-1})^{f}_{\, \ft'} \Bht^{\ft'}
\label{temp+1}
\end{eqnarray}
where we have dropped the expectation value signs and will
for the remainder of this section.
Taking the trace of this equation results in the identity
\begin{equation}
b^{f} \Bh_{f} = \bt_{\ft} \Bht^{\ft}.
\label{one+1}
\end{equation}
Inserting this back into (\ref{temp+1}) leads to
\begin{equation}
b^{f} \Bh_{f'} (m^{-1})^{f'}_{\, \ft} = 
\bt_{\ft} \Bht^{\ft'} (m^{-1})^{f}_{\, \ft'}.
\label{two+1}
\end{equation}
This equation is enough to imply that
\begin{eqnarray}
\Bh_{f} & = & \bt_{\ft} m^{\ft}_{\, f} \xi \nonumber \\
\Bht^{\ft} & = & m^{\ft}_{\, f} b^f \xi
\label{three+1}
\end{eqnarray}
where we have introduced the unknown $\xi$.
One further result that we need from (\ref{temp+1}) can
be obtained by contracting it with either $\Bh_f$ or $\Bht^{\ft}$.
After some simplification and use of the relations
(\ref{three+1}), we find  
\begin{eqnarray}
(m^{-1})^{f}_{\, \ft} \Bht^{\ft}
& = & \frac{1}{ S} \Mh^{f}_{\, \ft} \Bht^{\ft} \nonumber \\
\Bh_{f} (m^{-1})^{f}_{\, \ft} & = & \frac{1}{ S}
\Bh_{f} \Mh^{f}_{\, \ft}.
\label{four+1}
\end{eqnarray}
Using (\ref{three+1}) and (\ref{four+1}) we may rewrite
(\ref{temp+1}) as
\begin{equation}
m^{\ft}_{\, f} = ( S - \frac{\Bh_g \Mh^{g}_{\, \gt} \Bht^{\gt}}{ S \xi})
(\Mh^{-1})^{\ft}_{\, f} - \frac{1}{ S \xi} \Bht^{\ft} \Bh_{f}.
\label{five+1}
\end{equation}
Combined with (\ref{three+1}) we have managed to express the Lagrange
multipliers in terms of the desired set of variables up to the 
unknown $\xi$.

To solve for $\xi$ we must now bring in the equations (\ref{second+1}).
Eliminating $m^{\ft}_{\, f}$ and $b^g$ or $\bt_{\gt}$ via the
relations (\ref{five+1}) and (\ref{three+1}) respectively results
after a bit of algebra in a linear equation for $\xi$ with solution
\begin{equation}
\xi = - \frac{\det \Mh - \Bh_{g} \Mh^{g}_{\, \gt} \Bht^{\gt}}{S^2}.
\label{xi+1}
\end{equation}
This finally allows us to express the Lagrange multipliers in
terms of the desired variables.  We can now integrate the 
equations (replacing the operators $\Mh$ etc. by their
vevs) for $W_{eff}$ (\ref{dWone}) to find
\begin{equation}
W_{eff} = - S \ln \Bigl( \frac{\det M - 
B_{g} M^{g}_{\, \gt} \Bt^{\gt}}{\Lambda^{2N}} \Bigr)
+ f(S)
\end{equation}
up to the $S$ dependent function $f(S)$.  $f(S)$ is given
by the Veneziano-Yankielowicz term (\ref{WVY}) for
$N_f = N_c +1$.  Substituting and integrating out $S$ we find for the 
low energy effective superpotential
\begin{equation}
W_{eff} = - \frac{1}{\Lambda^{2N-3}} 
(\det M -B_{g} M^{g}_{\, \gt} \Bt^{\gt} )
\end{equation}
in agreement with the results of \cite{Seiberg:1994bz}.

\section{$ N_f > N_c + 1$}

The results obtained thus far are not new, only the 
methods to obtain them are somewhat novel.  The
more interesting cases in which to apply these methods
are when the number of flavors is even larger, i.e.,
when $N_f > N_c + 1$.  In this case one can hope
to address questions about confinement, chiral
symmetry breaking, and Seiberg duality \cite{Seiberg:1995pq}.
While the methods used above are straightforward
to apply to the cases of $N_f > N_c + 1$, the
equations that one obtains are more complicated
and difficult to solve.  We shall outline the
construction in this section and leave its
possible solution to future work.

As before we consider a superpotential which consists
solely of Lagrange multiplier terms enforcing the 
meson and baryon vevs,
\begin{equation}
W = m^{\ft}_{\, f} ( \Mh^{f}_{\, \ft} - M^{f}_{\, \ft})
+ b^{f_1 \cdots f_{\Nt}} 
( \Bh_{f_1 \cdots f_{\Nt}} - B_{f_1 \cdots f_{\Nt}}) + 
\bt_{\ft_1 \cdots \ft_{\Nt}} (\Bht^{\ft_1 \cdots \ft_{\Nt}} - 
\Bt^{\ft_1 \cdots \ft_{\Nt}})
\label{supermany}
\end{equation}
where we have defined $\Nt := N_f - N_c = N_f - N$ and the
baryonic operators
\begin{eqnarray}
\Bh_{f_1 \cdots f_{\Nt}} & := &  \frac{1}{N !} 
\epsilon_{f_1 \cdots f_{\Nt} g_1 \cdots g_N}
\epsilon^{a_1 \cdots a_N} Q^{g_1}_{a_1} \cdots Q^{g_N}_{a_N} \\
\Bht^{\ft_1 \cdots \ft_{\Nt}} & := & \frac{1}{N !} 
\epsilon^{\ft_1 \cdots \ft_{\Nt} \gt_1 \cdots \gt_N}
\epsilon_{a_1 \cdots a_N} \Qt_{\gt_1}^{a_1} \cdots \Qt_{\gt_N}^{a_N}.
\end{eqnarray}
The remaining fields in the superpotential, $m^{\ft}_{\, f}$,
$b^{f_1 \cdots f_{\Nt}}$ and $\bt_{\ft_1 \cdots \ft_{\Nt}}$ are
the Lagrange multipliers.

As before we first construct the Konishi anomaly equations
which are given by
\begin{eqnarray}
\langle \Mh^{f}_{\, \ft} \rangle m^{\ft}_{\, f'} & = & 
( \langle S \rangle - b^{f_1 \cdots f_{\Nt}}
 \langle \Bh_{f_1 \cdots f_{\Nt}} \rangle ) \delta^{f}_{f'}
+ \Nt  b^{f_1 \cdots f_{\Nt -1} f'} 
\langle \Bh_{f_1 \cdots f_{\Nt -1} f} \rangle \nonumber \\
m^{\ft'}_{\, f} \langle \Mh^{f}_{\, \ft} \rangle  & = & 
( \langle S \rangle - \bt_{\ft_1 \cdots \ft_{\Nt} } 
\langle \Bht^{\ft_1 \cdots \ft_{\Nt}} \rangle) 
\delta^{\ft'}_{\ft} + \bt_{\ft_1 \cdots \ft_{\Nt -1} \ft'} 
\langle \Bht^{\ft_1 \cdots \ft_{\Nt -1} \ft} \rangle.
\label{first+many}
\end{eqnarray}
The other pair of equations needed to solve for the low
energy effective action follow in direct analogy to 
before.  We define the operators
\begin{eqnarray}
\Bht_{a}^{\ft_1 \cdots \ft_{\Nt +1}} & := & \frac{1}{N!} 
\epsilon^{\ft_1 \cdots \ft_{\Nt + 1} \gt_2 \cdots \gt_N}
\epsilon_{a a_2 \cdots a_N} \Qt_{\gt_2}^{a_2} \cdots \Qt_{g_N}^{a_N}
\nonumber \\
\Bh^{a}_{f_1 \cdots f_{\Nt +1}} & := & \frac{1}{N!} 
\epsilon_{f_1 \cdots f_{\Nt + 1} g_2 \cdots g_N}
\epsilon^{a a_2 \cdots a_N} Q^{g_2}_{a_2} \cdots Q^{g_N}_{a_N}.
\end{eqnarray}
We note that these transform under the $SU(N_c)$ symmetry exactly
as $Q^{f}_{a}$ and $\Qt^{a}_{\ft}$ respectively and therefore
for vanishing superpotential one has the exact conservation
laws 
\begin{eqnarray}
{\bar D}^2 (Q^{\dagger}_{f} e^V \Bht^{\ft_1 \cdots \ft_{\Nt +1}} ) & = & 0
\nonumber \\
{\bar D}^2 (\Bh_{f_1 \cdots f_{\Nt +1}} e^V \Qt^{\ft \, \dagger} ) 
& = & 0.
\end{eqnarray}
Putting the superpotential (\ref{supermany}) back in and taking
expectation values results in the equations
\begin{eqnarray}
(N & - & 1)! m^{\ft}_{\, f} (\langle \Bht^{\gt_1 \cdots \gt_{\Nt}} \rangle 
\delta^{\gt}_{\ft} - \sum_{j=1}^{\Nt} 
\langle \Bht^{\gt_1 \cdots \gt_{j-1} \gt \gt_{j+1} \cdots \gt_{\Nt}}
\rangle
\delta^{\gt_j}_{\ft} ) \nonumber \\
& + & b^{f_1 \cdots f_{\Nt}}
\epsilon_{f_1 \cdots f_{\Nt} p_1 \cdots p_{N-1} f}
\epsilon^{\gt_1 \cdots \gt_{\Nt} \qt_1 \cdots \qt_{N-1} \gt}
\langle \Mh^{p_1}_{\, \qt_1} \rangle
 \cdots \langle  \Mh^{p_{N-1}}_{\, \qt_{N-1}} \rangle = 0 \nonumber \\
(N & - & 1)! m^{\ft}_{\, f} (
\langle \Bh_{g_1 \cdots g_{\Nt}} \rangle 
\delta_{g}^{f} - \sum_{j=1}^{\Nt} 
\langle
\Bh_{g_1 \cdots g_{j-1} g g_{j+1} \cdots g_{\Nt}} \rangle
\delta_{g_j}^{f} ) \nonumber \\
 & + & \bt_{\ft_1 \cdots \ft_{\Nt}}
\epsilon^{\ft_1 \cdots \ft_{\Nt} \pt_1 \cdots \pt_{N-1} \ft}
\epsilon_{g_1 \cdots g_{\Nt} q_1 \cdots q_{N-1} g}
\langle \Mh^{q_1}_{\, \pt_1} \rangle  \cdots 
\langle \Mh^{q_{N-1}}_{\, \pt_{N-1}} \rangle = 0
\label{second+many}
\end{eqnarray}
where we have used the factorization property of gauge invariant,
chiral operators in writing these equations.

The procedure for solving these equations for the Lagrange
multipliers in terms of the meson and baryon expectation
values is in principle straightforward.  We begin by
multiplying (\ref{second+many}) by $(\Nt + 1)$ $\Mh$ operators
and contracting with the free upper/lower indices respectively.
This produces a $\det \Mh$ operator in the second term.
Next we use (\ref{first+many}) to eliminate the $m^{\ft}_{\, f}$
Lagrange multiplier in (\ref{second+many}).
After some algebra we arrive at
\begin{eqnarray}
\bt_{\gt_1 \cdots \gt_{\Nt}} & = & - \frac{1}{\Nt ! \, \det \Mh}
\Bigg( \Bigl( S + (\frac{N_f}{N} - 2)(b \cdot \Bh) \Bigr)
(\Bh \Mh)_{\gt_1 \cdots \gt_{\Nt}} \nonumber \\
& - & \Bigl( \frac{N_f}{N} - 1
\Bigr) \sum_{j=1}^{\Nt} (b \cdot \Bh)^{f}_{\, f'}
(\Mh^{-1})^{\ft}_{\, f} \Mh^{f'}_{\gt_j}
(\Bh \Mh)_{\gt_1 \cdots \gt_{j-1} \ft \gt_{j+1} \cdots \gt_{\Nt}}
\Bigg) \nonumber \\
b^{g_1 \cdots g_{\Nt}} & = & - \frac{1}{\Nt ! \, \det \Mh}
\Bigg( \Bigl( S + (\frac{N_f}{N} - 2)(\bt \cdot \Bht) \Bigr)
(\Mh \Bht)^{g_1 \cdots g_{\Nt}} \nonumber \\
& - & \Bigl( \frac{N_f}{N} - 1
\Bigr) \sum_{j=1}^{\Nt} (\bt \cdot \Bht)_{\ft}^{\, \ft'}
(\Mh^{-1})^{\ft}_{\, f} \Mh^{g_j}_{\ft'}
(\Mh \Bht)^{g_1 \cdots g_{j-1} f g_{j+1} \cdots g_{\Nt}}
\Bigg)
\label{bbtmany}
\end{eqnarray}
where we have introduced some notation to simplify the 
expression,
\begin{eqnarray}
(\Bh \Mh)_{\gt_1 \cdots \gt_{\Nt}} & := & \Bh_{g_1 \cdots g_{\Nt}}
\Mh^{g_1}_{\, \gt_1} \cdots \Mh^{g_{\Nt}}_{\, \gt_{\Nt}}
\nonumber \\
(b \cdot \Bh)^{f}_{\, f'} & := & b^{g_1 \cdots g_{\Nt -1} f}
\Bh_{g_1 \cdots g_{\Nt -1} f'} \nonumber \\
(b \cdot \Bh) & := & b^{g_1 \cdots g_{\Nt}} \Bh_{g_1 \cdots g_{\Nt}}
\end{eqnarray}
with the other quantities defined similarly.

At this stage we can eliminate one Lagrange multiplier,
$b^{g_1 \cdots g_{\Nt}}$ say, to find a linear equation for
$\bt_{\gt_1 \cdots \gt_{\Nt}}$ in terms of the meson
and baryon expectation values.  Unlike the previous
cases however the resulting equation is considerably
more difficult to solve.  The key difference 
is that the last terms on the right-hand-sides
of the equations in (\ref{bbtmany}) were of the same
form as the first terms on the right-hand-sides.  In
that case solving the equations was trivial because they
could be reduced in the end to solving for a single 
variable.  In this case no such reduction (as far as
we know) is possible making solving the equations more
involved.

One interesting point to note however is that the solution
for the Lagrange multipliers is necessarily linear in the
gluino condensate $S$.  This is easy to see by simply
scaling all Lagrange multipliers in (\ref{bbtmany}) by
$S$ and noting that $S$ then factors out of the equation.
The same holds true in the equations for $m^{\ft}_{\, f}$
in (\ref{first+many}).  As a consequence, after adding
in the Veneziano-Yankielowicz contribution, the low
energy effective superpotential will take the form
\begin{equation}
W_{eff} = (N_c - N_f) S \Bigl( - \ln \frac{S}{\Lambda^3} + 1 \Bigr)
+ S f(M,B,\Bt)
\end{equation}
where $f(M,B,\Bt)$ denotes the contribution coming from solving
the equations above.

\section{Seiberg dual theories}

When the number of flavors satisfies $N_c + 2 \leq N_f \leq
(3/2) N_c$, the moduli space of the gauge theory discussed
in the previous section is known to be singular at the
point $M^{f}_{\, \ft} = B_{f_1 \cdots f_{\Nt}} = 
\Bt^{\ft_1 \cdots \ft_{\Nt}} = 0$.  It was conjectured
by Seiberg that the singularity is due to the presence
of particles becoming massless at this point in moduli
space.  Seiberg conjectured a dual IR description
of this theory which moreover is valid at the origin
of moduli space.  This theory consists of $N_f$
chiral matter fields $q_{f}$ and $\qt^{\ft}$, 
a set of chiral scalar fields $\Mh^{f}_{\ft}$,
and superpotential $W_m = \Mh^{f}_{\ft} \, \qt^{\ft}
q_{f}$ where the gauge group is $SU(N_f - N_c)$ and the
gauge indices have been suppressed.

There are several non-trivial consistency checks that
are discussed in detail in \cite{Seiberg:1994bz} that
this dual description must pass in order to have a chance
to describe the same IR physics as the theory discussed
in the previous section.  First note that the global
symmetries are the same.  Secondly we note that the massless
degrees of freedom away from the point 
$\Mh^{f}_{\, \ft} = 0$ are the same.
To see this let us first define the dual mesonic and
baryonic operators
\begin{eqnarray}
\Nh^{\ft}_{\, f} & := & \qt^{\ft} q_f \nonumber \\
\Bh^{(m)}_{f_1 \cdots f_{\Nt}} & := & \epsilon_{a_1 \cdots a_{\Nt}}
q^{a_1}_{f_1} \cdots q^{a_{\Nt}}_{f_{\Nt}} \nonumber \\
\Bht_{(m)}^{\ft_1 \cdots \ft_{\Nt}} & := & \epsilon^{a_1 \cdots a_{\Nt}}
\qt_{a_1}^{\ft_1} \cdots \qt_{a_{\Nt}}^{\ft_{\Nt}}
\end{eqnarray}
where the $(m)$ super or subscript denotes that this
is the dual baryonic operator.  Because of the superpotential
the dual meson is massive away from the point
$\Mh^{f}_{\ft} = 0$.  The baryonic operators
of the dual theories have identical transformation properties,
identifying them then the dual theories in the end are described by
the same set of low energy degrees of freedom.
Finally one can check that the 't Hooft anomaly matching conditions
are satisfied.

To go even farther one would like to compute the low energy 
effective superpotential on both sides of the duality and
show that they are the same.  While we will be unable to 
do this we nevertheless think that it will be useful
to use the analysis described above to find a set of 
equations whose solution would provide the low energy
effective superpotential.  The analysis is essentially
identical to above, the only new features here are that
in this case there is already a superpotential present,
aside from the Lagrange multiplier terms that we will
add, and also that this time around we have an extra
set of scalar chiral matter fields to deal with which
leads to more equations.

The superpotential that we will take is
\begin{eqnarray}
W_{m} & = & \Mh^{f}_{\, \ft} \Nh^{\ft}_{\, f} +
m^{\ft}_{\, f} ( \Mh^{f}_{\, \ft} - M^{f}_{\, \ft})
+ b^{f_1 \cdots f_{\Nt}} 
( \Bhm_{f_1 \cdots f_{\Nt}} - \Bm_{f_1 \cdots f_{\Nt}}) \nonumber \\
& + & 
\bt_{\ft_1 \cdots \ft_{\Nt}} (\Bhtm^{\ft_1 \cdots \ft_{\Nt}} - 
\Btm^{\ft_1 \cdots \ft_{\Nt}})
\label{supermag}
\end{eqnarray}
where as before the Lagrange multipliers $m^{\ft}_{\, f}$,
$b^{f_1 \cdots f_{\Nt}}$ and $\bt_{\ft_1 \cdots \ft_{\Nt}}$
enforce the vev constraints.  This form is valid provided
that the vev of the scalar fields $\Mh^{f}_{\, \ft}$ is 
non-zero, which is all that we shall concern ourselves with
here.  The analysis now proceeds as before.  In complete
analogy to the previous sections (essentially one just
replaces $Q$'s with $q$'s) we find the set of Konishi anomaly
equations (dropping expectation value signs as before)
\begin{eqnarray}
\Mh^{f}_{\, \ft} \Nh^{\ft}_{\, f'} & = & S \delta^{f}_{f'}
- \Nt b^{f_1 \cdots f_{\Nt - 1} f} \Bhm_{f_1 \cdots f_{\Nt - 1} f'}
\nonumber \\
\Mh^{f}_{\, \ft} \Nh^{\ft'}_{\, f} & = & S \delta^{\ft'}_{\ft}
- \Nt \bt_{\ft_1 \cdots \ft_{\Nt - 1} \ft} 
\Bhtm^{\ft_1 \cdots \ft_{\Nt - 1} \ft'} 
\label{first+mag}
\end{eqnarray}
and the set of conservation equations
\begin{eqnarray}
\Mh^{f}_{\, \ft} \Bhtm^{\ft_1 \cdots \ft_{\Nt - 1} \ft}
& = & - \Nt ! \, b^{f_1 \cdots f_{\Nt - 1} f}
\Nh^{\ft_1}_{\, f_1} \cdots 
\Nh^{\ft_{\Nt - 1}}_{\, f_{\Nt - 1}} \nonumber \\
\Mh^{f}_{\, \ft} \Bhm_{f_1 \cdots f_{\Nt - 1} f}
& = & - \Nt ! \, \bt_{\ft_1 \cdots \ft_{\Nt - 1} \ft}
\Nh^{\ft_1}_{\, f_1} \cdots 
\Nh^{\ft_{\Nt - 1}}_{\, f_{\Nt - 1}}.
\label{second+mag}
\end{eqnarray}
Finally we can get another set of equations by recalling
that $\Mh^{f}_{\, \ft}$ is a chiral, gauge invariant scalar
field.  Taking the expectation value of its equation of motion
leads to 
\begin{eqnarray}
m^{\ft}_{\, f} + \Nh^{\ft}_{\, f} = 0.
\label{third+mag}
\end{eqnarray}

The strategy before was to solve for the Lagrange multipliers
as functions of the low energy degrees of freedom
$\Mh^{f}_{\, \ft}$, $\Bhm_{f_1 \cdots f_{\Nt}}$ and
$\Bhtm^{\ft_1 \cdots \ft_{\Nt}}$.  This leads to
a set of equations in complete analogy to 
(\ref{dWone}) which can in principle be integrated
to find $W_{eff}$ up to an $S$-dependent term.
As an example, we can easily use the equations above
to construct an equation for $\bt_{\ft_1 \cdots \ft_{\Nt}}$.
Multipy the second equation of (\ref{second+mag})
by an $\Mh$ on every free index.  Then use (\ref{third+mag})
and (\ref{first+mag}) to replace the $\Nh$ dependence in 
terms of the desired variables.  We find in the end
\begin{eqnarray}
0 & = & \Mh^{f_1}_{\, \ft^{'}_1} \cdots \Mh^{f_{\Nt}}_{\, \ft^{'}_{\Nt}}
\Bhm_{f_1 \cdots f_{\Nt}} \nonumber \\
& + & \Nt ! \, 
\bt_{\ft_1 \cdots \ft_{\Nt - 1} \ft_{\Nt}^{'}}
( S \delta^{\ft_1}_{\ft_{1}^{'}} - \Nt 
(\bt \cdot \Bhtm)_{\ft_{1}^{'}}^{\, \ft_1}) \cdots 
( S \delta^{\ft_{\Nt - 1}}_{\ft_{\Nt - 1}^{'}} - \Nt 
(\bt \cdot \Bhtm)_{\ft_{\Nt - 1}^{'}}^{\, \ft_{\Nt -1}}),
\label{solvebtilde}
\end{eqnarray}
i.e., a nonlinear equation for $\bt$ in contrast to the
linear equation that we obtained previously (\ref{bbtmany})
for the IR dual theory.

It is interesting to note that some information
on the $S$-dependence of
$W_{eff}$ can be obtained without solving for
$W_{eff}$ explicitly.  If we scale the Lagrange
multipliers $b$ and $\bt$ as well as 
the mesons $\Mh^{f}_{\, \ft}$ by $S$, then
all $S$ dependence divides out of (\ref{solvebtilde})
as well as from (\ref{first+mag}).  Consequently
we know that $W_{eff}$ must take the form, up to
an $S$-dependent constant of integration,
$W_{eff} = S g(M/S,B,\Bt)$ for some function $g$.
Putting in the Veneziano-Yankielowicz contribution
we find
\begin{equation}
W_{eff} = - N_c S \Big(- \ln \frac{S}{\Lambda^3} + 1 \Big)
+ S g(M/S,B,\Bt).
\end{equation}

\section{Matrix model approach}

In light of the conjecture of Dijkgraaf and Vafa, it is
natural to ask if the results obtained earlier can
be obtained from a matrix model.  
This question has
been discussed by various authors 
\cite{Argurio:2002hk,Bena:2002ua,Suzuki:2002jc,Bena:2003vk}.
Recall that the
Dijkgraaf-Vafa conjecture relates the planar diagrammatric
contribution to the matrix model free energy to
the low energy effective superpotential of the gauge
theory.  Moreover the matrix model action is given
by the gauge theory superpotential expressed in 
terms of the elementary superfields.  The question
becomes somewhat more subtle when baryons are involved,
for two reasons.  The first is that normally one extracts planar
diagrams from the matrix model by taking a large
$N$ limit, that is, the planar diagrammatic 
contribution dominates in this limit with all
planar diagrams scaling with $N$ in the same way.
When baryons are included, the diagrams that
one would like to call ``planar''\footnote{Some generalization
of the definition of planar is needed here due to
the presence of the $\epsilon$ tensor in the definition
of a baryon, see eg. \cite{Bena:2002ua}.} no longer 
all scale with $N$ in the same
way, making their extraction complicated.  A second
reason that baryons are more troublesome here is
that the previous discussion assumed that the
number of flavors $N_f$ was also taken large, i.e.
$N_f = N + k$ where $k$ is fixed, but it has been
shown that when fundamental matter is included, only planar 
diagrams with at most a single boundary are relevant
to the low energy effective superpotential.  Taking
$N_f$ large will not help to isolate the planar
diagrams with different numbers of boundaries. 
Nevertheless, it has been shown in \cite{Bena:2002ua} that the
generalized planar diagrams (in the gauge theory) are the relevant ones
that give rise in the end to the low energy effective
superpotential.  If it is also true that the generalized
planar diagrams from the associated matrix model
are relevant to computing $W_{eff}$, then extracting them 
remains another question.

In this section we derive a set of matrix model loop equations
for arbitrary $N_c$ and $N_f \geq N_c$.  These equations
have identical form to those constructed in previous sections
provided that we identify the glueball field $S$ with
$g_m /N$ where $g_m$ is the matrix model coupling constant,
except that we don't have the factorization property
that followed in the gauge theory case from chirality.
Normally on the matrix model side this property follows
from the large $N$ limit, namely in the large $N$ limit
the disconnected diagrams dominate the correlator giving
rise to factorization.  If one can make sense of the large
$N$ limit in this case, then this might be good enough
to derive the factored form of the loop equations thereby
giving identical results to the gauge theory analysis above.
However the naive large $N$ limit does not produce sensible
results as the different terms in the loop equation scale
with $N$ differently.

Deriving the matrix model loop equations is a straightforward
matter.  As an example we shall consider the $N_c = N_f$
case.  Our matrix model is given by the integral
\begin{equation}
e^{- N^2 {\cal F}} = \int \, [dQ] \, [d \Qt] e^{-(N/g_m) S}
\end{equation} 
with action $S$ given by the superpotential (\ref{superzero}) 
considered previously in the $N_f = N_c$ case.  Here
the matrix model variables $Q^{f}_a$ and $\Qt^{a}_{\ft}$
are $N \times N$ where $N_c = N_f = N$.  To derive the loop equations we
consider first the identity
\begin{equation}
0=\int \, [dQ] \, [d \Qt] \frac{\partial}{\partial Q^{f}_{a}}
\Bigg( Q^{f'}_{a} \, e^{-(N/g_m) S} \Bigg).
\label{firstloop}
\end{equation}
Expanding this out results in the equation
\begin{equation}
\langle \Mh^{f}_{\, \ft} \rangle m^{\ft}_{\, f'}  = 
( \langle S \rangle - b \langle \Bh \rangle ) \delta^{f}_{f'}
\end{equation}
where we have identified $S = g_m / N$.
Replacing the explicit $Q$ variables by $\Qt$ variables in
(\ref{firstloop}) results in another loop equation
of the same form in terms of the $\bt$ and $\Bht$ fields.
These loop equations are in fact identical to those in the
gauge theory case.

The second set of loop equations is obtained by considering
the identity
\begin{equation}
0=\int \, [dQ] \, [d \Qt] \frac{\partial}{\partial Q^{f}_{a}}
\Bigg( \frac{1}{N!} \epsilon_{a a_2 \cdots a_N}
\epsilon^{\ft \ft_2 \cdots \ft_N} \Qt^{a_2}_{\ft_2} \cdots
\Qt^{a_N}_{\ft_N} \, e^{-(N/g_m) S} \Bigg).
\end{equation}
Expanding out the derivative results in 
\begin{equation}
m^{\ft}_{\, f} \langle \Bht \rangle + 
b \langle \det \Mh   ( \Mh^{-1})^{\ft}_{\, f} \rangle  =  0.
\end{equation}
Exchanging the $Q$ and $\Qt$ variables results in a similar
type loop equation involving $\bt$ and $\Bht$.  The important 
point to note here is that
the second term is an expectation value of a product of 
operators.  To reproduce the gauge theory results we must
replace this with a product of expectation values of individual
mesons.

As we said before, this factorization is usually accomplished
by taking a large $N$ limit, however the different
terms in the above loop equations scale differently 
with $N$ making this limit less than helpful.  Perhaps
there is a double scaling limit that one could take that
preserves the form of the loop equations while giving
rise to factorization.

\bigskip
\centerline{\bf Acknowledgements}
\noindent
We thank D.A. Lowe for comments and especially
C-I. Tan for numerous discussions.
This research is supported in part by DOE
grant DE-FE0291ER40688-Task A.

\bibliography{dvpubs}
\bibliographystyle{h-elsevier}
\end{document}